\documentclass{jpsj3}
\usepackage{color}

\def\gsim{\mathop {\vtop {\ialign {##\crcr 
$\hfil \displaystyle {>}\hfil $\crcr \noalign {\kern1pt \nointerlineskip } 
$\,\sim$ \crcr \noalign {\kern1pt}}}}\limits}
\def\lsim{\mathop {\vtop {\ialign {##\crcr 
$\hfil \displaystyle {<}\hfil $\crcr \noalign {\kern1pt \nointerlineskip } 
$\,\,\sim$ \crcr \noalign {\kern1pt}}}}\limits}

\title
{Unconventional Quantum Criticality due to Critical Valence Transition}

\author{
Kazumasa {\sc Miyake}$^{\,1}$ and Shinji {\sc Watanabe}$^{\,2}$
} 

\inst
{$^{1}$Toyota Physical and Chemical Research Institute, Nagakute, Aichi 480-1192, Japan\\
$^{2}$ Department of Basic Sciences, Kyushu Institute of Technology, Kitakyushu, 804-8550, Japan
}

\recdate
{
December 7, 2013
}

\abst
{
Quantum criticality due to the valence transition in some Yb-based heavy fermion metals has gradually 
turned out to play a crucial role to understand the non-Fermi liquid properties that cannot 
be understood from the conventional quantum criticality theory due to magnetic transitions. 
Namely, critical exponents giving the temperature ($T$) dependence of the resistivity $\rho(T)$, 
the Sommerfeld coefficient, $C(T)/T$, the  magnetic susceptibility, $\chi(T)$, and 
the NMR relaxation rates, $1/(T_{1}T)$, can be understood as the effect of the critical valence 
fluctuations of $f$ electrons in Yb ion in a unified way.  
There also exist a series of Ce-based heavy fermion metals 
that exhibit anomalies in physical quantities, enhancements of the residual resistivity $\rho_{0}$ 
and the superconducting critical temperature ($T_{\rm c}$) around the pressure where 
the valence of Ce sharply changes.  Here we review the present status of these problems 
both from experimental and theoretical aspects.  
}
\kword
{
Unconventional quantum criticality, critical valence transition, non-Fermi liquid, heavy fermions
}

\begin{document}
\maketitle

\section{Introduction}
Since the mid 1990's, the physics of quantum critical phenomena has been intensively discussed 
in the heavy fermion community. This is reasonable because the heavy fermions usually arise 
in nearly magnetic materials, in which strong local correlation due to the strong local Coulomb repulsion 
of the order of 1 Ryd ($\simeq$13.6~eV) works among $f$-electrons at rare earth or actinide ions.  
For example, 
CeCu$_2$Si$_2$, the first heavy fermion superconductor discovered by Steglich {\it et al}. 
in 1979,~\cite{Steglich1} 
has now turned out to be located close to the quantum critical point where the spin density wave state 
disappears.~\cite{Stockert} A series of compounds were reported to exhibit magnetic quantum 
critical point under pressure around which an unconventional superconductivity 
appears: e.g., in CePd$_2$Si$_2$~\cite{Grosche}, CeIn$_3$~\cite{Mathur}, and 
CeRh$_2$Si$_2$~\cite{Movshovich}.  
As for the mechanism of superconductivity, the ``antiferromagnetic" spin fluctuation 
mechanism was proposed in mid 1980's.~\cite{MSRV,SLH}  After that, its strong coupling treatments 
have been developed~\cite{Monthoux} and also for a mechanism of high-$T_{\rm c}$ 
cuprates and some organic superconductors.~\cite{MoriyaUeda}  

On the other hand, an importance of valence fluctuations was already suggested as an origin for 
an enhancement of the superconducting (SC) transition temperature $T_{\rm c}$ of CeCu$_2$Si$_2$ under 
pressure by Bellarbi {\it et al}. in 1984,~\cite{Bellarbi} at relatively early stage of research 
in heavy fermion superconductivity.  The Kondo-volume-collapse mechanism for CeCu$_2$Si$_2$ 
proposed by Razafimandaby {\it et al}. at the same stage is also based on a kind of valence 
fluctuation idea.~\cite{Razafi} In SCES conference at Paris in 1998, Jaccard reported 
comprehensive data on pressure-induced superconductivity in CeCu$_2$Ge$_2$, 
a sister compound of CeCu$_2$Si$_2$, together with anomalous properties in its normal 
state,~\cite{Jaccard} in which a close relationship between the enhancement 
of $T_{\rm c}$ and a sharp valence crossover of Ce ion from Kondo to valence-fluctuation regime 
was explicitly shown. Theoretical attempts to coherently understand these experimental results  
have been performed since then.~\cite{MNO, Onishi1, Onishi2, MM, Holmes, WIM} 
Similar detailed measurements, including specific heat measurement under pressure, 
in CeCu$_2$Si$_2$ was also reported by Holmes {\it et al}.,~\cite{Holmes} 
on the basis of almost local critical valence 
fluctuation scenario.  A remarkable report on CeCu$_2$(Si$_{0.9}$Ge$_{0.1}$)$_2$ by 
Yuan {\it et al}.~\cite{Yuan} also eloquently indicated the existence of the valence 
fluctuation mechanism other than that due to critical ``antiferromagnetic" fluctuations, because 
the $T_{\rm c}$ exhibits two domes, one at the magnetic quantum critical pressure and another at 
the pressure where the valence of Ce ion changes sharply as in the case of CeCu$_2$Si$_2$ and 
CeCu$_2$Ge$_2$. 

These findings show that only the so-called  Doniach phase diagram,~\cite{Doniach} 
a sort of {\it dogma} in heavy fermion physics, is not sufficient to fully understand 
the physics of heavy fermions{. In other words,} the Kondo lattice model, 
in which the $f$ electron 
number $n_{\rm f}$ per ion is fixed  as $n_{\rm f}=1$, would not be a sufficient model but the Anderson 
lattice model offers us a better starting point.  A prototypical valence transition 
phenomenon is a $\gamma$-$\alpha$ transition of Ce metal that exhibits the first order 
valence transition at $P \simeq 1.0$ GPa from Ce$^{+3.03}$ to Ce$^{+3.14}$ at $T=300$K and 
has a critical end point (CEP) at  $T \simeq 600$K and  $P \simeq 2.0$GPa.~\cite{AlphaGamma} 
There were roughly two ways to understand the valence transition in rare earth ions: 
Kondo volume collapse (KVC) model and extended periodic Anderson model (PAM) or 
extended Falicov-Kimball model (FKM). 

The KVC model { uses the fact that the Kondo temperature $T_{\rm K}$, 
representing the energy gain due to the Kondo effect or correlation, has a sharp 
pressure (volume) dependence through the c-f exchange interaction.  Then, 
the valence transition is discussed} 
in terms of the Gibbs free energy.~\cite{Coqblin,Allen,Dzero}  Although it describes 
quite well a $P$-$V$ or $P$-$T$ phase diagram of the $\gamma$-$\alpha$ transition of Ce metal 
and other Ce- or Yb-based heavy fermions, the criticality is not {\it directly} related with  
the valence change nor to the response of electron degrees of 
freedom but with volume or strain of the crystal{. Therefore,} the relation between microscopic 
model and phenomena is not straightforward as far as we understand.  For example, 
it seems not so simple to understand possible magnetic anomalies near the quantum critical end point 
(QCEP) of valence transition, such as anomalous {temperature ($T$)} dependences 
in magnetic susceptibility and NMR/NQR relaxation rates. (See \S3.)  

\begin{figure}[h]
\begin{center}
\rotatebox{0}{\includegraphics[width=0.5\linewidth]{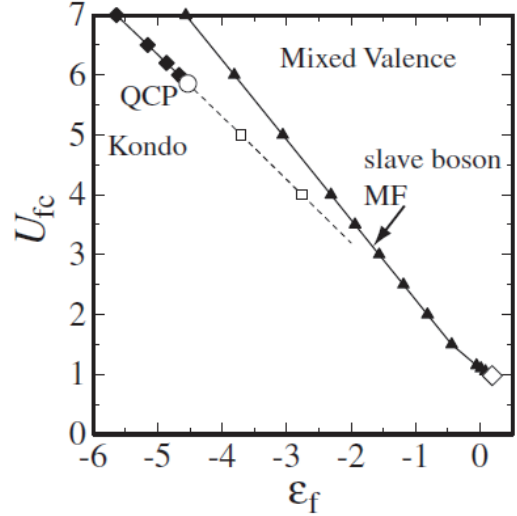}}
\caption{
Phase diagram at $T=0$ of the system described by the Hamiltonian (\ref{eq:1}) in 
$\varepsilon_{\rm f}-U_{\rm fc}$ 
plane. Triangles ({diamonds and} squares) are the results by 
the slave-boson mean-field approximation 
(DMRG calculations 
{for the one dimensional (1d) version of the Hamiltonian (\ref{eq:1})}
). Solid lines represent the first order valence transition, and dashed line 
represent that for the valence crossover from Kondo to {mixed valence} regime.  Closed circles 
are critical end point of the first order valence 
transition{, i.e., quantum crtical point of valence transition}.  Parameters are $V/t=0.1$, 
$U_{\rm ff}/t=100$, $t$ being the transfer integral of tight-binding model for the conduction band
{in the 1d version of the Hamiltonian (\ref{eq:1})}, 
and the total electron number per unit cell, $n$, is fixed as $n=7/8$, the same as 
used in ref.\ \citen{Onishi1}. Unit of $U_{\rm fc}$ and $\varepsilon_{\rm f}$ are also $t$.  
}
\label{Fig:1}
\end{center}
\end{figure}

On the other hand, the FKM directly discusses the valence state of rare earth ions 
by considering the condition how it is influenced by the effect of the Coulomb 
repulsion $U_{\rm fc}$ between $f$- and conduction electrons.~\cite{FalicovKimball,Varma1}  
However, original FKM includes no c-f hybridization which is the heart of the valence 
fluctuation problem including the Kondo effect.  After that, theoretical efforts 
have been performed to take into account 
the hybridization effect in a form or another in the case of lattice 
systems~\cite{HewsonRiseborough, Schlottmann} or as the impurity 
problem.~\cite{CostiHewson, TakayamaSakai, Perakis, Khomskii}  
The Hamiltonian of the extended {PAM or extended FKM} is given as 
\begin{eqnarray}
H_{\rm EPAM}
 &=& \sum_{{\bf k} \sigma}(\epsilon_{\bf k}-\mu) c_{{\bf k} \sigma}^{\dagger}c_{{\bf k} \sigma}
 +\varepsilon_{\rm f} \sum_{{\bf k} \sigma}f_{{\bf k} \sigma}^{\dagger}f_{{\bf k} \sigma} 
 +U_{\rm ff}\sum_i n_{i \uparrow}^{\rm f} n_{i \downarrow}^{\rm f}%
\nonumber \\
 & &\qquad+V\sum_{{\bf k} \sigma}(c_{{\bf k} \sigma}^{\dagger}f_{{\bf k} \sigma}^{}+{\rm h.c.})
 +U_{\rm fc}\sum_{i \sigma \sigma'}n_{i \sigma}^{\rm f}n_{i \sigma'}^{\rm c},
 \label{eq:1}
\end{eqnarray}
where $U_{\rm fc}$, the f-c Coulomb repulsion, is included other than the
conventional PAM. Here, it should be noted that $\sigma$ in eq.\ (\ref{eq:1}) stands for the 
label specifying the degrees of freedoms of the Kramers doublet state 
of the ground crystalline-electric-field (CEF) level.   
If there is no hybridization, $V=0$, the condition of valence transition is 
given by 
\begin{equation}
\varepsilon_{\rm f}+n_{\rm c}U_{\rm fc}=\mu,
\label{eq:2}
\end{equation}
where $\mu$ is the chemical potential or the Fermi level 
{in the Kondo limit where $f$ electrons are essentially singly occupied. Even in the 
case} $V\not=0$, this condition is valid in the mean-field level of approximation. 
Figure\ \ref{Fig:1} shows the ground state phase diagrams in 
the $\varepsilon_{\rm f}$-$U_{\rm fc}$ plane that is obtained  
by the mean-field approximation using the slave boson technique by taking into account 
the strong correlation effect ($U_{\rm ff}=\infty$), and by the 
density-matrix-renormalization-group (DMRG) method for the one-dimensional version 
of the Hamiltonian eq.\ (\ref{eq:1}).~\cite{WIM}  Calculations on the Gutzwiller variational 
ansatz have also been performed.~\cite{Onishi2,Saiga, Sugibayashi,Kubo,Hagymasi} 
The first order valence transition line is given essentially by condition (\ref{eq:2}).  
The QCEP (closed circles) for the first order valence transition (line) shifts from the position 
given by the mean-field approximation to that given by asymptotically exact DMRG calculation 
due to the strong quantum fluctuation effect.  In this approach based on the extended FKM or 
extended PAM, properties of electronic state associated with the valence change or critical 
fluctuations can be {\it directly} calculated within a required accuracy as shown in \S3. 

Another important issue is how to understand the unconventional quantum critical phenomena observed 
in a series of materials,~\cite{Bauer,Seuring1,Steglich2,Ishida,Loehneysen,Nakatsuji,Matsumoto,Deguchi}
in which the critical exponent of $T$ dependence in 
various physical quantities cannot be understood from the conventional quantum criticality 
theory associated with magnetic transition.~\cite{Moriya, MoriyaTakimoto,Hertz,Millis}  
Indeed, Table\ \ref{Table:1} shows the $T$-dependence (at low temperatures) of the resistivity $\rho(T)$, 
the Sommerfeld coefficient $C(T)/T$, uniform magnetic susceptibility $\chi(T)$, and the NMR/NQR 
relaxation rates $1/{(T_{1}T)}$ 
together with predictions for these quantities by the theory for 
three-dimensional antiferromagnetic quantum critical point (QCP) and those for the QCEP of valence 
transition.  It is clear, as shown 
in Table\ \ref{Table:1}, that the critical exponents observed are totally different from those 
of conventional ones near the antiferromagnetic QCP, but agree with those given by the theory 
of the critical valence fluctuations (CVF) that gives the critical exponent $\zeta$ as 
$0.5\lsim \zeta\lsim 0.7$ depending on the region of $T$.~\cite{WM:PRL,WM:JPCM}  

\begin{table}
\caption{Theoretical results for conventional criticality due to antiferromagnetic (AF) QCP 
for a series of physical quantities, and unconventional criticality for those observed 
in a series of materials together with theoretical result due to critical valence 
fluctuations (CVF) giving the exponent $\zeta$ as $0.5\lsim \zeta \lsim<0.7$ depending on the region temperature 
$T$ higher than $T_{0}$, the extremely small temperature scale (see \S3).  
The symbol * indicates that there is no available experiment. \\
\ 
}
\label{Table:1}
\begin{center}
{\offinterlineskip
\halign{\strut\vrule#&\quad#\hfil\quad&
              \vrule#&\quad#\hfil\quad&
              \vrule#&\quad#\hfil\quad&
              \vrule#&\quad#\hfil\quad&
              \vrule#&\quad#\hfil\quad&
              \vrule#&\quad#\hfil\quad&
              \vrule#\cr
      \noalign{\hrule}
&Theories \&  Materials&& 
  $\rho(T)$ && $C(T)/T$ && $\chi(T)$&& $1/T_{1}T$&&
Refs. &\cr
      \noalign{\hrule}
& AF QCP && 
  $T^{3/2}$ && ${\rm const.}-T^{1/2}$ && ${\rm const.}-T^{1/4}$ && $T^{-3/4}$ &&
 \ \ \citen{MoriyaTakimoto,Hatatani}&\cr
      \noalign{\hrule}
& YbRh$_2$Si$_2$ && 
  $T$ && $-\log\,T$ && $T^{-0.6}$ && $T^{-0.5}$&&
 \ \ \citen{Steglich2,Ishida}&\cr
      \noalign{\hrule}
& $\beta$-YbAlB$_4$ &&  $T^{1.5}\sim T$ &&
 $-\log\,T$ && $T^{-0.5}$ && * &&
 \ \ \citen{Nakatsuji} &\cr
      \noalign{\hrule}
& YbCu$_{3.5}$Al$_{1.5}$&& {$T^{1.5}\sim T$} && 
 $-\log\,T$ && $T^{-0.66}$ && * &&
 \ \ \citen{Bauer,Seuring1}&\cr
      \noalign{\hrule}
& Yb$_{15}$Au$_{51}$Al$_{31}$ &&  
  $T$ && $-\log\,T$ && $T^{-0.51}$ && $T^{-0.51}$&&
 \ \ \citen{Deguchi}&\cr
      \noalign{\hrule}
& CVF && 
  $T$ && $-\log\,T$ && $T^{-\zeta}$ && $T^{-\zeta}$ &&
 \ \ \citen{WM:PRL,WM:JPCM}&\cr
      \noalign{\hrule}
}
}
\end{center}
\end{table}

In order to understand this unconventional quantum criticality, several scenarios such as the local 
criticality theory on the so-called Kondo breakdown idea,~\cite{Si,Coleman,Si2} 
a theory of the tricritical point,~\cite{Misawa} 
a theory based on the 
model specific to $\beta$-YbAlB$_4$,~\cite{Coleman2} and so on, have been proposed so far.  
While these theories appear to have succeeded in explaining a certain part of anomalous behaviors of this 
criticality, their success seems to remain partial one to our knowledge. 
On the other hand, we have recently 
developed a theory based on the CVF near the QCEP of valence 
transition,~\cite{WM:PRL,WM:JPCM} explaining the exponents shown in Table\ 1 in a unified way. 

The purpose of the present paper is to review theoretical and experimental status of 
the unconventional criticality 
based on CVF together with its background that reinforces a solidity of this idea. 
Organization of the paper is as follows.  
In \S2, we present a series of experimental facts in some Ce-based heavy fermion metals,  
offering us a persuasive experimental evidence for a reality of sharp crossover in 
the valence of Ce ion under high pressures.  Cases of CeCu$_2$(Si,Ge)$_2$ and CeRhIn$_5$ 
are discussed, together with 
theoretical developments that enable us to understand these salient experimental facts in a 
unified way from the point of view of CVF. 
In \S3, it is discussed how a mode-mode coupling theory for the CVF is constructed on the 
basis of the Hamiltonian (\ref{eq:1}) in parallel to the case of the mode-mode coupling theory 
for magnetic fluctuations starting with the Hubbard model.~\cite{MoriyaTakimoto,Hertz, Millis}  
The critical exponents of temperature dependence given by the theory is shown to explain those 
exponents listed in Table\ \ref{Table:1} quite well.  
In \S4, remaining problems and prospect of CVF is discussed. One is related with a reality of 
the model Hamiltonian (\ref{eq:1}) which gives lager valence change than that observed 
experimentally, and another is concerned with the effect of excited CEF levels of Ce ion. 

\section{Reality of Sharp {Valence Crossover} in Ce-Based Heavy Fermions}
In this section, we briefly summarize the present status and discuss to what extent 
the reality of sharp crossover in valence of Ce ion in Ce-based heavy fermions, 
CeCu$_2$(Si,Ge)$_2$ and CeRhIn$_5$.  

\subsection{{\rm CeCu}$_2${\rm (Si,Ge)}$_2$}
Here we present a series of experimental evidence of sharp crossover of Ce valence under 
pressure in CeCu$_2$Si$_2$,~\cite{Holmes} while it was first reported~\cite{Jaccard} and 
discussed~\cite{MNO} for CeCu$_2$Ge$_2$.  

\begin{figure}[h]
\begin{center}
\rotatebox{0}{\includegraphics[width=0.5\linewidth]{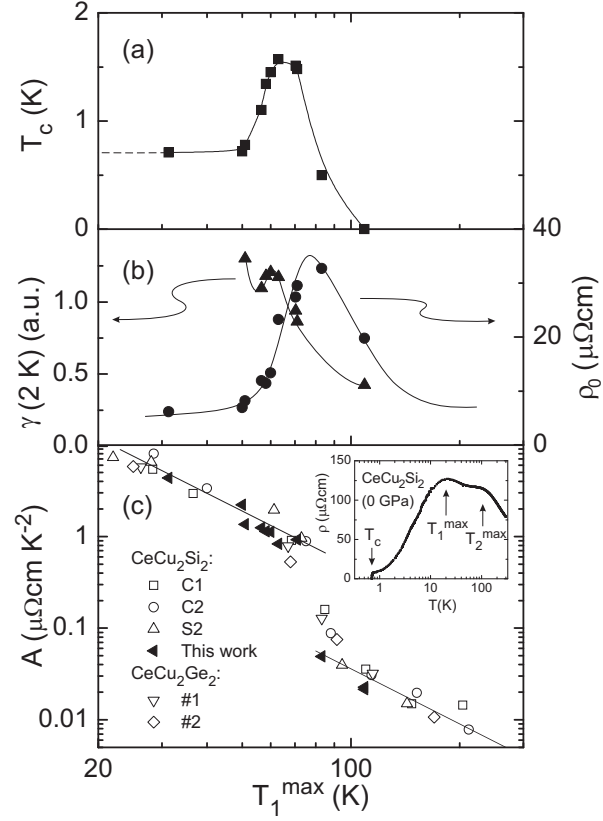}}
\caption{
Plotted against $T_1^{\rm max}$ (defined in inset),
     a measure of the characteristic energy scale of the system,
     are (a) the bulk superconducting transition temperature, (b) the
     residual resistivity and $\gamma$ coefficient of the
     electronic specific heat, and (c) the coefficient $A$ of the
     $\rho\sim A T^2$ law of resistivity.  Note the straight lines
     where the expected $A\propto (T_1^{\rm max})^{-2}$ scaling is
     followed.  The maximum of $T_{\rm c}$ coincides with the start of
     the region where the scaling relation is broken, while the
     maximum in residual resistivity is situated in the middle of
     the collapse in $A$.  Pressure increases towards the right-hand side
     of the scale (high $T_{\rm K}$). \cite{Holmes} 
}
\label{Fig:2}
\end{center}
\end{figure}

Most direct signature of sharp valence crossover is 
a drastic decrease of the $A$ coefficient of the $T^{2}$ resistivity law
by about two orders of magnitude around the pressure $P=P_{\rm v}$ where the residual 
resistivity $\rho_{0}$ exhibits a sharp and pronounced peak as shown in 
Figs.\ \ref{Fig:2}(a)$\sim$(c).  
Since $A$ scales as $(m^{*})^{2}$ in the so-called Kondo regime, this implies
that the effective mass $m^{*}$ of the quasiparticles also
decreases sharply there.  This fall of $m^{*}$ is a direct signature a sharp change of 
valence of Ce, deviating from Ce$^{3+}$, since the following approximate (but canonical) 
formula holds in the strongly correlated limit:\cite{RiceUeda,Shiba}
\begin{equation}
\label{eq:3m*nf}
{m^{*}\over m_{\rm band}} ={1-n_{\rm f}/2\over 1-n_{\rm f}},
\end{equation}
where $m_{\rm band}$ is the band mass without electron correlations, and 
$n_{\rm f}$ is the f-electron number per Ce ion.  

This sharp crossover of the valence 
is consistent with a sharp crossover of the so-called Kadowaki-Woods (KW) ratio,\cite{rf:KW}
$A/\gamma^{2}$, where $\gamma$ is the Sommerfeld coefficient of
the electronic specific heat, from that of a strongly correlated class to a weakly correlated 
one. $\gamma^{-1}$ can be identified with the Kondo temperature $T_{\rm K}$,
which is experimentally accessible by resistivity measurements as $T_{1}^{\rm max}$ 
shown in the inset of Fig.\ \ref{Fig:2}(c).  
This indicates that the mass enhancement due to the dynamical electron correlation is quickly 
lost at around $P\sim P_{\rm v}$.~\cite{rf:MMV}

\begin{figure}
\begin{center}
\includegraphics[width=0.7\linewidth]{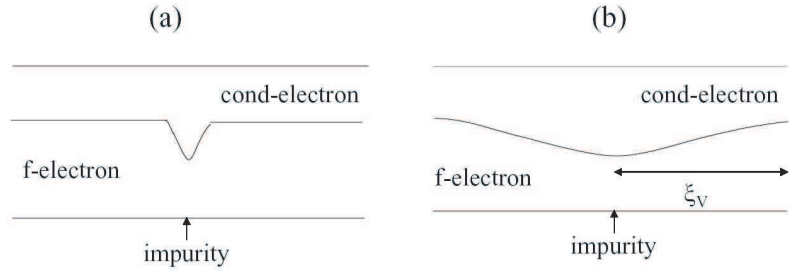}
\caption{
Schematic view of charge distribution of f- and conduction electrons 
around impurity: (a) at far from $P\sim P_{\rm v}$ where the effect of 
impurity remains as short-ranged so that the residual resistivity $\rho_{0}$ 
is not enhanced; 
(b) at around $P\sim P_{\rm v}$ where the effect of impurity extends  
to long-range region, because the correlation length $\xi_{\rm v}$ of 
valence fluctuations diverges as $P\rightarrow P_{\rm v}$, 
leading to highly enhanced $\rho_{0}$.  
}
\label{Fig:3}
\end{center}
\end{figure}

The huge peak of  $\rho_{0}$ at around $P\sim P_{\rm v}$ can be understood as a many-body 
effect enhancing the impurity potential. In the forward scattering limit, 
this enhancement is proportional to the valence susceptibility 
$-(\partial n_{\rm f}/\partial \varepsilon_{\rm f})_{\mu}$, where 
$\varepsilon_{\rm f}$ is the atomic f-level of the Ce ion, and $\mu$ 
is the chemical potential~\cite{MM}.  Physically speaking, 
local valence change coupled to the impurity or disorder gives 
rise to the change of valence in a wide region around the impurity 
which then scatters the quasiparticles quite strongly, leading to 
the increase of $\rho_{0}$ (see Fig.\ \ref{Fig:3}).
On the other hand, the effect of AF critical fluctuations 
on $\rho_{0}$ is rather moderate as discussed in ref.\ \citen{MN}.  
Thus, the critical pressure $P_{\rm v}$ can be clearly defined by 
the maximum of $\rho_0$.  

Other characteristic behaviors shown in Fig.\ \ref{Fig:2} near $P=P_{\rm v}$ are 
peak of the $T_{\rm c}$ and the Sommerfeld coefficient $\gamma(T=2~{\rm K})$ 
at slightly lower pressure than $P_{\rm v}$.  These behaviors can be understood on the basis of 
explicit theoretical calculations in which almost local valence fluctuations of Ce is shown to 
develop around the pressure 
where the sharp valence crossover occurs.~\cite{Onishi1,WIM,Holmes, Miyake:JPCM} 
Another characteristic behavior is that the $T$-linear behavior is observed in $[\rho(T)-\rho_{0}]$ 
over rather wide temperature range above $T_{\rm c}$. This also can be understood on the 
basis of a picture of the almost local CVF,~\cite{Holmes} 
which is related to the issue discussed in \S3. 

\begin{figure}
\begin{center}
\includegraphics[width=0.6\linewidth]{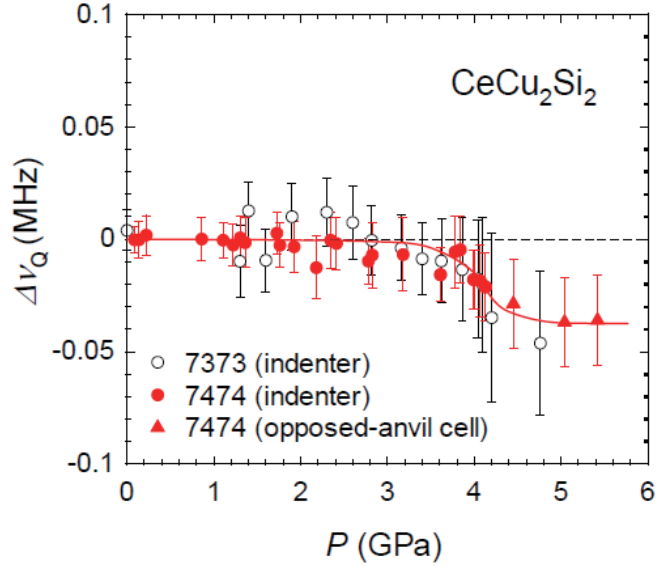}
\caption{(Color online) 
Pressure dependence of deviation from the linear $P$ dependence of the 
NQR frequency $^{63}\nu_{\rm Q}$ {in CeCu$_2$Si$_2$.~\cite{Kobayashi}}   
}
\label{Fig:4}
\end{center}
\end{figure}

Much more direct evidence for the sharp crossover of the valence of Ce ion in CeCu$_2$Si$_2$ 
was obtained by $^{63}$Cu-NQR measurements at temperature down to $T=3.1\,$K and under pressures 
up to $P=5.5\,$GPa passing 
$P_{\rm v}\simeq 4.5\,$GPa.~\cite{Fujiwara,Fujiwara2,Kobayashi}  Namely, the NQR frequency 
$^{63}\nu_{\rm Q}$ suddenly deviates at above 4$\,$GPa from the linear $P$-dependence in the 
low pressure range ($P\le 3.5\,$GPa). This sudden downward deviation of $^{63}\nu_{\rm Q}$ 
can be regarded as due to an increase of Ce valence, because the linear $P$-dependence 
is recovered again at $P>4.5\,$GPa.  The $P$-dependence of the deviation in $^{63}\nu_{\rm Q}$ is 
shown in Fig.\ \ref{Fig:4}.  Corresponding change of the valence $\Delta n_{\rm f}$ was estimated 
to be $\Delta n_{\rm f}=0.05$ by the first principles calculations,~\cite{Kobayashi} 
which may give the change in $n_{\rm f}$, say from $n_{\rm f}=0.99$ to $n_{\rm f}=0.94$ corresponding 
to that of decrease of mass enhancement by one order of magnitude (two orders of magnitude 
in the $A$ coefficient of the resistivity). 

It should be mentioned, however, that measurements of the X-ray powder diffraction (XRPD) at $T=12\,$K  
under presser (in CeCu$_2$Si$_2$) detected no sudden change in variations of lattice constant 
except for a very tiny change (of the same order as the experimental resolution)  
at around $P=4.5\,$GPa.~\cite{Kobayashi}  
In ref.\ \citen{Kobayashi}, it was also reported that the result of similar measurements in 
CeCu$_2$Ge$_2$ shows no detectable change in the volume at around $P=15\,$GPa in contrast to the 
report of ref.\ \citen{Onodera}.  The difference in {two} results was suggested to be due to that of 
pressure medium.  Recent measurements of the X-ray absorption spectroscopy at Ce L3 edge 
at $T=14\,$K under high pressure in CeCu$_2$Si$_2$ also shows no 
discontinuous change at around $P=4.5\,$GPa,~\cite{Rueff} which is not inconsistent with the result 
of XRPD in ref.\ \citen{Kobayashi}. Origin of this discrepancy on valence change by two probes, 
NQR and X-ray, is {not clear} for the moment.

\subsection{{\rm CeRhIn}$_5$}
\begin{figure}
\begin{center}
\includegraphics[width=0.8\linewidth]{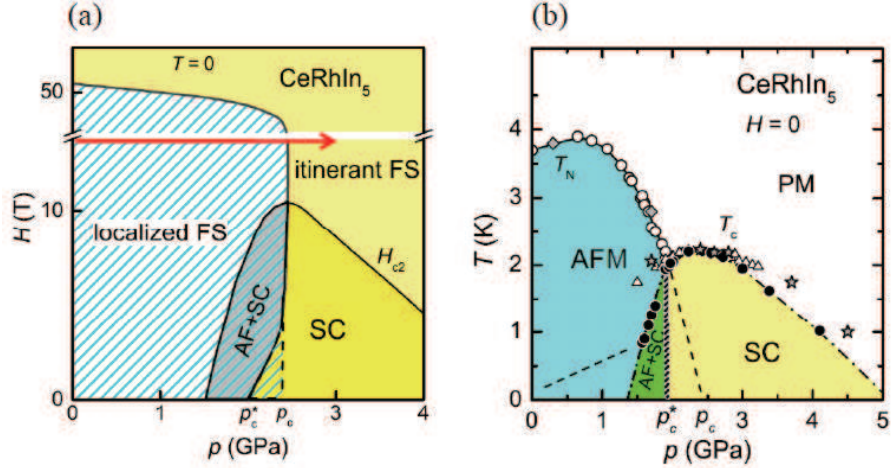}
\caption{(Color online) 
(a) Phase diagram of CeRhIn$_5$ at $T\to 0$~K in $P$-$H$ plane.~\cite{Knebel} 
(b) Phase diagram of CeRhIn$_5$ without magnetic field ($H=0$~Tesla) in $P$-$T$ plane.~\cite{Knebel}
The dashed line indicates the superconducting transition temperature reported in ref.\ \citen{Chen}. 
}
\label{Fig:5}
\end{center}
\end{figure}

CeRhIn$_5$ is a prototypical heavy fermion system in which superconductivity and antiferromagntic 
order coexist under pressure.~\cite{Knebel,Park} The phase diagram is shown in Fig.\ \ref{Fig:5}: 
(a) in the $P$-$H$ plane at $T=0\,$K, and (b) in the $P$-$T$ plane at $H=0$.  
Measurements of de Haas-van Alphen (dHvA) effect have been performed along the arrow in 
Fig.\ \ref{Fig:5}(a),~\cite{Shishido} revealing the following aspects: (1) the Fermi surfaces change at 
$P=P_{\rm c}$ from those expected for localized $f$ electrons (as in LaRhIn$_5$) to 
those for itinerant $f$ electrons. (2) the cyclotron mass exhibits a sharp peak at 
around $P=P_{\rm c}$.  At $P=P_{\rm c}$, the AF order disappear{s suddenly}, 
corresponding to the 
discontinuous change of the Fermi surfaces. It is mysterious that the effective mass of quasiparticles 
increases steeply towards $P=P_{\rm c}$ where the first order magnetic transition occurs.  

\begin{figure}
\begin{center}
\includegraphics[width=0.8\linewidth]{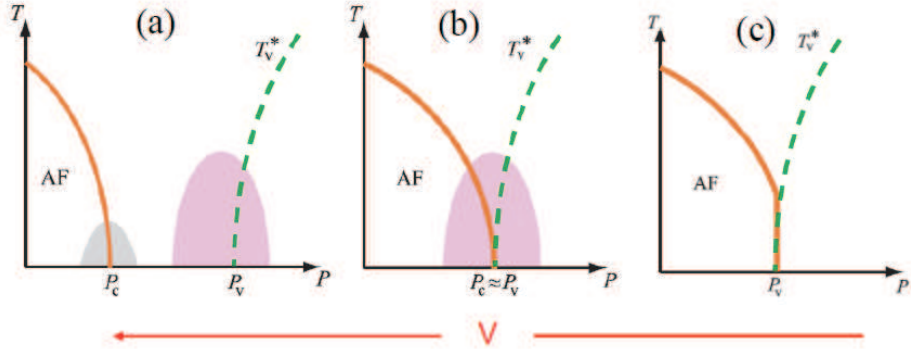}
\caption{(Color online) 
Schematic $P$-$T$ Phase diagrams in $P$-$T$ plane for three cases corresponding to different strength 
of the hybridization $V$. 
}
\label{Fig:6}
\end{center}
\end{figure}

This problem was resolved by a theoretical analysis on the basis of the extended PAM, eq.\ (\ref{eq:1}), 
supplemented by the Zeeman term,  
\begin{equation}
H_{\rm mag}=-h\sum_{i}(S^{\rm f}_{iz}+S^{\rm c}_{iz}),
\label{eq:4}
\end{equation}
where $h\equiv g\mu_{\rm B}H$.  This Hamiltonian was treated by the mean-filed approximations both 
for the AF order and the slave boson which is introduced to take into account 
the strong local correlation effect between on-site $f$ electrons.~\cite{WM:JPSJ,WM:JPCM2}  
Depending on the strength of the hybridization $V$ in the Hamiltonian, eq.\ (\ref{eq:1}), qualitative 
phase diagrams in the $P$-$T$ plane change as shown in Fig.\ \ref{Fig:6}. In the systems with relatively 
large $V$, the AF phase is suppressed so that $P_{\rm c}$, corrsponding to 
the AF-QCP, and $P_{\rm v}$, coresponding to the QCEP of valence transition or 
valence crossover, is well separated, as in CeCu$_2$(Si,Ge)$_2$ and Ce(Co,Ir)In$_5$ 
(Fig.\ \ref{Fig:6}(a)).  
In the systems with relatively weak $V$, the region of AF state extends to higher pressure region 
so that the {\it intrinsic} $P_{\rm c}$ becomes lager than  $P_{\rm v}$.  
However, the AF order is cut by the 
valence crossover at $P=P_{\rm v}$ where AF order vanishes discontinuously, as in 
CeRhIn$_5$ (Fig.\ \ref{Fig:6}(c)).  
There occurs the case where $P_{\rm c}$ coincides with $P_{\rm v}$ at a certain strength of 
$V$ (Fig.\ \ref{Fig:6}(b)), giving a possible new type of quantum critical phenomena 
as in YbRh$_2$Si$_2$~\cite{Steglich2}.  

The results of microscopic calculations under the magnetic field $H$ for the case Fig.\ \ref{Fig:6}(c) 
are summarized as follows:~\cite{WM:JPSJ}
\begin{enumerate}
\item 
Both lines of the first order valence transition and valence crossover almost coincides with 
that of AF transition of the first order; i.e., AF order is cut by the valence transition or 
valence crossover.~{\cite{Comment}}  
\item 
Associated with the first order transition, the Fermi surface 
changes discontinuously from smaller size to larger size corresponding to the transition  
from a so-called ``localized" $f$ electrons to ``itinerant" ones.  
\item 
The effective mass of quasiparticles exhibits a sharp peak structure around $P=P_{\rm c}$ as 
the band effect of folding or unfolding of the Fermi surface associated with the AF transition.  
\item 
The effective mass of quasiparticles is already enhanced in the AF state from that given by the 
first-principles band structure calculations, implying that the hybridization between $f$ and 
conduction electrons is not vanishing there at all.  
\end{enumerate}
These results capture the essential experimental aspects of CeRhIn$_5$ obtained by dHvA experiments 
in ref.\ \citen{Shishido}

\begin{figure}
\begin{center}
\includegraphics[width=0.5\linewidth]{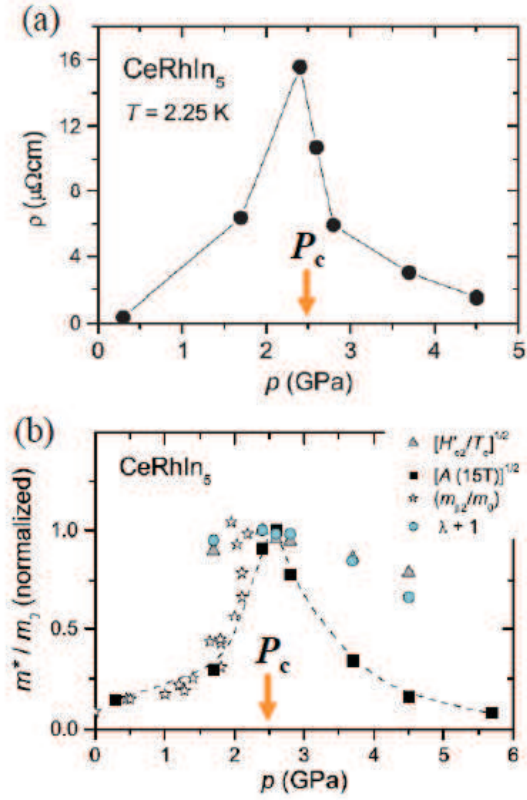}
\caption{(Color online) 
(a) {E}nhanced resistivity at $T=T_{\rm c}$, (b) scaling of $\sqrt{A}$ and $m^{*}$ 
{in CeRhIn$_5$}.     
}
\label{Fig:7}
\end{center}
\end{figure}

Other experimental evidence for the valence crossover to be realized in CeRhIn$_5$ at $P=P_{\rm c}$ 
are the following three{:  
\begin{enumerate}
\item[a)]
The resistivity at $T=2.25\,$K just above $T_{\rm c}$ exhibits 
huge peak at $P=P_{\rm c}$~\cite{Knebel,Muramatsu} 
as in the case of CeCu$_2$(Si,Ge)$_2$ as shown in Fig.\ \ref{Fig:7}(a). 
This strongly suggests that the valence fluctuations are growing sharply around 
$P=P_{\rm c}$ as discussed in ref.\ \citen{MM}.   
\item[b)] 
The exponent of $\alpha$, representing the $T$-dependence of 
$[\rho(T)-\rho_{0}]\propto T^{\alpha}$ approach $\alpha=1$ near $P=P_{\rm c}$ as 
demonstrated by Park {\it et al}. in ref.\ \citen{Park}.   
This is also the signature of critical valence fluctuations.~\cite{Holmes} 
\item[c)] 
The Kadowaki-Wood scaling, $\sqrt{A}/m^{*}={\rm const}.$ or $A/\gamma^{2}={\rm const}.$, 
holds at $P\lsim P_{\rm c}$, although both $A$ and $m^{*}$ grow steeply as $P$ approach 
$P_{\rm c}$ from lower pressure side as shown in Fig.\ \ref{Fig:7}(b).~\cite{Knebel}  
This behavior cannot be understood from the scenario based on the AF criticality where 
$A/\gamma^{2}$ diverges.  
\end{enumerate}
}

Finally, we note that the pressure dependence of the SC transition temperature $T_{\rm c}$ 
and the upper critical field $H_{{\rm c}2}$ are quite different (see Fig.\ \ref{Fig:5}(a)).  
Namely, the former is almost 
flat at $P>P_{\rm c}$ while the latter prominently increases as $P_{\rm c}$ is approached 
from the higher pressure side. This suggests that the SC pairing interaction is promoted 
by the magnetic field $H$ itself.  One of such possibility is that the QCEP of valence 
transition is located at the magnetic field $H=H^{*}$ ($>10\,$Tesla) on the phase boundary 
between AF and normal state in the phase diagram Fig.\ \ref{Fig:5}(a).  This is not so ridiculous 
idea, considering that its phase boundary coincides with the valence crossover lines as mentioned 
in the item (1) above,~\cite{WM:JPSJ} and the SC state is stabilized in the region where 
a sharp crossover of valence occurs.~\cite{Onishi1,WIM,Holmes,Miyake:JPCM}

\section{Mode-Mode Coupling Theory for Critical Valence Fluctuations}
In this section, we outline the theory for the critical exponents due to the 
critical valence fluctuations (CVF) shown in Table\ \ref{Table:1}.~\cite{WM:PRL} 
We start with the Hamiltonian (\ref{eq:1}) and construct the mode-mode coupling 
theory in parallel to the case of the theory for magnetic QCP which starts 
with the Hubbard model.~\cite{Moriya,Hertz}  

\subsection{Formalism}
In the model Hamiltonian (\ref{eq:1}), the on-site Coulomb repulsion $U_{\rm ff}$ 
between $f$ electrons is the strongest interaction, so that we first take into account 
its effect and after that construct the mode-mode coupling theory for critical valence 
fluctuations caused by the $f$-$c$ repulsion $U_{\rm fc}$.  To consider the correlation effect 
due to $U_{\rm ff}$, we introduce the slave-boson operator $b_{i}$ 
to eliminate the doubly-occupied state, representing the effect of $U_{\rm ff}\to\infty$, 
under the constraint 
\begin{equation}
\sum_{m}n^{\rm f}_{im}+Nb^{\dagger}_{i}b_{i}=1,
\label{eq:6}
\end{equation}
where $i$ indicates the site of $f$ electron, and $m$ represents the generalized 
spin labels extended from $\sigma=\uparrow, \downarrow$ to  $m=1\cdots N$ for employing 
the large-$N$ expansion framework.~\cite{Onishi1}  

The Lagrangian is written as 
$
{\cal L}={\cal L}_{0}+{\cal L}'$: 
\begin{eqnarray}
& &{\cal L}_{0}=\sum_{{\bf k}m}c^{\dagger}_{{\bf k}m}
\left(
\partial_{\tau}+\bar{\varepsilon}_{\bf k}
\right)
c_{{\bf k}m}
+\sum_{{\bf k}{\bf k'}m}f^{\dagger}_{{\bf k}m}
\left(
\partial_{\tau}+\bar{\varepsilon}^{\rm f}_{{\bf k}-{\bf k'}}
\right)
f_{{\bf k'}m}
\nonumber
\\
& &\qquad\qquad
+\frac{V}{\sqrt{N_{\rm s}}}\sum_{{\bf k}{\bf k'}m}
\left(
c^{\dagger}_{{\bf k}m}f_{{\bf k'}m}b^{\dagger}_{{\bf k}-{\bf k'}}
+{\rm h.c.}
\right)
+\frac{N}{N_{\rm s}}\sum_{{\bf k}{\bf k'}}
b^{\dagger}_{\bf k}\lambda_{{\bf k}-{\bf k'}}b_{\bf k'}
\label{eq:7}
\\
& &{\cal L'}=
-\frac{U_{\rm fc}}{N}\sum_{im}
\left(
n^{\rm c}_{im}+n^{\rm f}_{im}
\right)
+\frac{U_{\rm fc}}{N}\sum_{imm'}
n^{\rm f}_{im}n^{\rm c}_{im'}, 
\label{eq:8}
\end{eqnarray}
where $N_{\rm s}$ is the number of lattice sites,  
$\lambda_{\bf k}$ is the Lagrange multiplier to impose the constraint, 
and 
$
\bar{\varepsilon}_{\bf k}\equiv\varepsilon_{\bf k}
+\frac{U_{\rm fc}}{N}
$ 
and 
$
\bar{\varepsilon}^{\rm f}_{{\bf k}-{\bf k'}}\equiv
\left(
\varepsilon_{\rm f}+\frac{U_{\rm fc}}{N}
\right)
\delta_{{\bf k}{\bf k'}}
+\frac{1}{\sqrt{N_{\rm s}}}
\lambda_{{\bf k}-{\bf k'}}
$. 
Here, we have separated $\cal L$ as ${\cal L}_{0}$ and ${\cal L'}$ 
to perform the expansion with respect to the $f$-$c$ Coulomb repulsion $U_{\rm fc}$.  

For the term $\exp(-S_{0})$ with the action 
$
S_{0}=\int_{0}^{\beta}d\tau{\cal L}_{0}(\tau)
$, 
the saddle point solution 
is obtained via the stationary condition $\delta S_{0}=0$ 
by approximating spatially uniform 
and time independent ones, i.e., $\lambda_{\bf q}=\lambda\delta_{\bf q}$ 
and $b_{\bf q}=b\delta_{\bf q}$. 
The solution is obtained by solving 
mean-field equations $\partial S_{0}/\partial\lambda=0$ 
and $\partial S_{0}/\partial b=0$ self-consistently.

To make the action 
$
S'=\int_{0}^{\beta}d\tau{\cal L'}(\tau), 
$ 
we introduce the identity applied by 
a Stratonovich-Hubbard transformation
\begin{equation}
{\rm e}^{-S'}
=\int{\cal D}\varphi\,
\exp
\left[
\sum_{im}
\int_{0}^{\beta}d\tau
\left\{
-\frac{U_{\rm fc}}{2}\varphi_{im}(\tau)^2
+{\rm i}\frac{U_{\rm fc}}{\sqrt{N}}
\left(
c_{im}f^{\dagger}_{im}-f_{im}c^{\dagger}_{im}
\right)
\varphi_{im}(\tau)
\right\}
\right].
\label{eq:9}
\end{equation}
Then, the partition function of the system is expressed as 
$
Z=\int{\cal D}(cc^{\dagger}ff^{\dagger}\varphi)\exp(-S)
$ 
with $S=S_{0}+S'$.  By performing Grassmann number integrals for $cc^{\dagger}$ 
and $ff^{\dagger}$, 
we obtain the action for the field $\varphi$ as (up to constant terms) 
\begin{eqnarray}
& &S\left[\varphi\right]=\sum_{m}\left[
\frac{1}{2}
\sum_{\bar{q}}\Omega_{2}(\bar{q})
\varphi_{m}(\bar{q})\varphi_{m}(-\bar{q})
\right.
\nonumber
\\
& & \qquad\qquad+
\left.
\sum_{\bar{q}_1,\bar{q}_2,\bar{q}_3}
\Omega_{3}(\bar{q}_1,\bar{q}_2,\bar{q}_3)
\varphi_{m}(\bar{q}_1)
\varphi_{m}(\bar{q}_2)
\varphi_{m}(\bar{q}_3)
\delta\left(\sum_{i=1}^{3}\bar{q}_{i}\right)
\right.
\nonumber \\
& & \qquad\qquad+
\left. 
\sum_{\bar{q}_1,\bar{q}_2,\bar{q}_3,\bar{q}_4}
\Omega_{4}(\bar{q}_1,\bar{q}_2,\bar{q}_3,\bar{q}_4)
\times
\varphi_{m}(\bar{q}_1)
\varphi_{m}(\bar{q}_2)
\varphi_{m}(\bar{q}_3)
\varphi_{m}(\bar{q}_4)
\delta\left(\sum_{i=1}^{4}\bar{q}_{i}\right)
+\cdots
\right]. 
\label{eq:10}
\end{eqnarray}
Here, a dominant part of the coefficient of the quadratic term is 
given by 
\begin{eqnarray}
\Omega_{2}({\bf q},i\omega_{l})=U_{\rm fc}
\left[1-\frac{2U_{\rm fc}}{N}
\chi^{\rm ffcc}_{0}({\bf q},i\omega_{l})
\right], 
\label{eq:11}
\end{eqnarray}
where
\begin{equation}
\chi^{\rm ffcc}_{0}({\bf q},i\omega_{l})
\equiv -\frac{T}{N_{\rm s}}\sum_{{\bf k},n}
G_{0}^{\rm ff}({\bf k}+{\bf q},i\varepsilon_{n}+i\omega_{l})
G_{0}^{\rm cc}({\bf k},i\varepsilon_{n}),
\label{eq:12}
\end{equation}
where $G_{0}^{\rm ff}({\bf k},{\rm i}\varepsilon_{n})$ and 
$G_{0}^{\rm cc}({\bf k},{\rm i}\varepsilon_{n})$ are the Green function 
of $f$ and conduction electrons for the saddle point solution for 
$U_{\rm ff}=\infty$, respectively.~\cite{WM:PRL}  

Since long wave length $|{\bf q}|\ll q_{\rm c}$ around ${\bf q}={\bf 0}$ 
and low frequency $|{\omega}|\ll \omega_{\rm c}$ regions play dominant roles 
in critical phenomena, with $q_{\rm c}$ and $\omega_{\rm c}$ being cutoffs 
for momentum and frequency which are of the order of inverse of the lattice constant 
and the effective Fermi energy, respectively.  Coefficients 
$\Omega_{i}$ for $i=2,3$, and 4 in eq.\ (\ref{eq:10}) are expanded for $q$ and $\omega$ around 
$({\bf 0},0)$ as follows:~\cite{WM:PRL} 
\begin{equation}
\Omega_{2}({\bf q},i\omega_{l})
\approx 
\eta_{0}+Aq^{2}+C_{q}\left|\omega_{l}\right|,
\label{eq:13}
\end{equation}
where 
\begin{equation}
\eta_{0} \equiv U_{\rm fc}
\left[1-\frac{2U_{\rm fc}}{N}
\chi^{\rm ffcc}_{0}({\bf 0},0)
\right],
\label{eq:14}
\end{equation}
and
\begin{equation}
\Omega_{3}(q_1,q_2,q_3)\approx \frac{v_3}{\sqrt{\beta N_{\rm s}}}, \quad
\Omega_{4}(q_1,q_2,q_3,q_4)\approx \frac{v_4}{\beta N_{\rm s}} .
\label{eq:15}
\end{equation}

{
We note that hereafter $A$ represents the coefficient of the 
$q^{2}$ term as in eq.\ (\ref{eq:13}), but not the coefficient of the $T^{2}$ term 
in the resistivity.
}

\subsection{Renormalization group analysis}
It is useful to analyze the property of the cubic and quartic terms in $\varphi$ in the 
action $S[\varphi]$, eq.\ (\ref{eq:10}), by the perturbation renormalization-group 
procedure:~\cite{Hertz} 
(a) Integrating out high momentum and frequency parts for $q_{\rm c}/s<q<q_{\rm c}$ and 
$\omega_{\rm c}/s^{z}<\omega<\omega_{\rm c}$, respectively, with 
$s$ being a dimensionless scaling parameter $(s\ge 1)$ and $z$ the dynamical exponent. 
(b) Scaling of $q$ and $\omega$ by $q'=sq$ and $\omega'=s^{z}\omega$. 
(c) Re-scaling of $\varphi$ by $\varphi'({\bf q'},\omega')=s^{a}\varphi({\bf q'}/s,\omega'/s)$. 
Then, we determine the scale factor $a$ so that the Gaussian term in eq.~(\ref{eq:10}) 
becomes scale invariant, leading to $a=-(d+z+2)/2$ with $d$ being spatial dimension. 
 Finally, the renormalization-group evolution for coupling constants 
$v_{j}$ ($j=3,\,4$) are given as  
\begin{eqnarray}
\frac{dv_{3}}{ds}&=&\left[6-(d+z)\right]v_{3}+{\cal O}(v_{3}^2), 
\label{eq:16}
\\ 
\frac{dv_{4}}{ds}&=&\left[4-(d+z)\right]v_{4}+O(v_{4}^2).
\label{eq:17}
\end{eqnarray}
By solving these equations, it is shown that higher order terms than 
the Gaussian term are irrelevant in the sense that 
\begin{eqnarray}
\lim_{s\to\infty}v_{j}(s)=0 \ \ {\rm for} \ \ j\ge 3
\label{eq:18}
\end{eqnarray}
for $d+z\ge 6$.  This implies that the upper critical dimension $d_{\rm u}$ for the cubic term 
to be irrelevant is $d_{\rm u}=6$.  
In the case of pure three dimensional system ($d=3$) exhibiting valence 
change uniform in space, where $C_{q}=C/q$, dynamical exponent $z$ is given by $z=3$, 
i.e., $d+z=6=d_{\rm u}$, so that the cubic term 
is marginally irrelevant~\cite{Miyake:JPCM}.  Hence, the universality class of the 
criticality of valence fluctuations belongs to the Gaussian fixed point. 
This implies that critical valence fluctuations are qualitatively described by the RPA 
framework with respect to $U_{\rm fc}$. 
The coefficient of the Gaussian term in eq.\ (\ref{eq:10}), i.e., eq.\ (\ref{eq:13}), 
is nothing but the inverse of the valence susceptibility 
$\Omega_{2}({\bf q},i\omega_{l})\equiv\chi_{\rm v}^{\rm RPA}({\bf q},i\omega_{l})^{-1}$.  
Namely, $\chi_{\rm v}^{\rm RPA}({\bf q},i\omega_{l})$ is given as 
\begin{equation}
\chi_{\rm v}^{\rm RPA}(q,i\omega_{l})=\frac{1}{\eta_{0}+Aq^2+C_{q}|\omega_{l}|}. 
\label{eq:19}
\end{equation}

However, there exist some cases with $z=2$ in general.  For example, if the effect of 
non-magnetic impurity scattering is taken into account, $C_{q}$ is given as 
$C_{q}=C/{\rm max}\{q,l_{\rm i}^{-1}\}$ with $l_{\rm i}$ being 
the mean free path of impurity scattering,~\cite{MNO} leading to $z=2$ unless the effect 
of impurity scattering is neglected.  
Another one is the case where a valence change occurs as a density wave with a finite 
ordered wave vector, giving the dynamical exponent $z=2$ in general. 
Nevertheless, the cubic coefficient vanishes on the 
line which is extending from the first-order transition line to the crossover region, as 
discussed by Landau in the case of gas-liquid transition.~\cite{Landau}
Namely, just at the QCEP of the valence transition, the cubic term is neglected safely, 
making the upper critical dimension of the system $d_{\rm u}=4$, but not $d_{\rm u}=6$, 
as far as the temperature dependence at the QCEP is concerned. 
Then, clean $(z=3)$ system and dirty $(z=2)$ system in three spatial dimension $(d=3)$  
are both above the upper critical dimension, i.e., $d+z>d_{\rm u}=4$. 
Thus, the higher order terms other than the Gaussian term are irrelevant in the action, 
which makes the fixed point Gaussian.

\subsection{Locality of CVF}
With the use of the saddle point solution for $G_{0}^{\rm ff}({\bf k},{\rm i}\varepsilon_{n})$ and 
$G_{0}^{\rm cc}({\bf k},{\rm i}\varepsilon_{n})$, 
we have found that the coefficient $A$ in eq.\ (\ref{eq:13}) or eq.\ (\ref{eq:19}) 
is extremely small of the order of $q_{\rm c}^{-2}{\cal O}(10^{-2}$), or 
almost dispersionless critical valence fluctuation mode appears near $q=0$ 
not only for deep $\varepsilon_{\rm f}$, i.e., in the Kondo regime, 
but also for shallow $\varepsilon_{\rm f}$, i.e., in the mixed valence regime,  
because of strong on-site Coulomb repulsion for $f$ electrons in the extended PAM, 
eq.\ (\ref{eq:1}).~\cite{WM:PRL}

The physical picture of emergence of this weak-$q$ dependence in the critical
valence fluctuation is analyzed as follows:~\cite{Miyake:JPCM}
The $q$-dependence in eq.\ (\ref{eq:13}) appears through 
$G_{0}^{\rm ff}({\bf k}+{\bf q},{\rm i}\varepsilon_{n})$ in 
$\chi_{0}^{\rm ffcc}({\bf q},,0)$, eq.\ (\ref{eq:12}).  
Near $q=0$, $\chi_{0}^{\rm ffcc}({\bf q},,0)$ is expanded as 
\begin{equation}
\chi_{0}^{\rm ffcc}(q,0)=\chi_{0}^{\rm ffcc}(0,0)
+\tilde{S}\left(\frac{V}{|\mu-\varepsilon_{\rm f}|}\right)^2q^2,
\label{eq:20}
\end{equation}
where $\tilde{S}$ includes the effect of the f-electron self-energy for $U_{\rm ff}$ 
in eq.\ (\ref{eq:1}). Since the f-electron self-energy has almost no $q$ dependence 
in heavy electron systems, the $q$  dependence of the $f$-electron propagator 
$G_{0}^{\rm ff}({\bf k}+{\bf q})$ comes from the hybridization $V$ with conduction 
electrons with the dispersion $\varepsilon_{{\bf k}+{\bf q}}$, as seen in the coefficient 
of the $q^2$ term in eq.\ (\ref{eq:20}).  Hence, the reduction of the coefficient $A$ in 
eq.\ (\ref{eq:19}) is caused by two factors. One is due to the smallness of 
$(V/|\mu-\varepsilon_{\rm f}|)^2$. In typical heavy electron systems, this factor is smaller 
than $10^{-1}$. The other one is the reduction of the coefficient $\tilde{S}$, which is suppressed 
by the effects of the on-site electron correlations $U_{\rm ff}$ in eq.\ (\ref{eq:1}). 
Numerical evaluations of $\chi_{0}^{\rm ffcc}(q,0)$ based on the saddle point solution for 
$U_{\rm ff}=\infty$ in eq.\ (\ref{eq:1}) show that extremely small $\tilde{S}$ appears not 
only in the Kondo regime, but also in the mixed-valence regime,~\cite{WM:PRL} 
indicating that the reduction by $\tilde{S}$ plays a major role. These multiple 
reductions are the reason why extremely small coefficient $A$ appears in eq.\ (\ref{eq:19}). 

The extremely small $A$ in eq.\ (\ref{eq:13}) or eq.\ (\ref{eq:19})  
makes the characteristic temperature for critical valence fluctuations
\begin{equation}
T_{0}\equiv\frac{Aq_{\rm B}^3}{2\pi C}
\label{eq:21}
\end{equation}
extremely small.  Here, $q_{\rm B}$ is a momentum at the Brillouin zone boundary. 
Hence, even at low enough temperature than the effective Fermi temperature 
of the system, i.e., so-called Kondo temperature, $T\ll T_{\rm K}$, 
the temperature scaled by $T_{0}$ can be very large: $t\equiv T/T_{0}\gg 1$. 
This is the main reason why unconventional criticality emerges at ``low" temperatures, 
which will be explained below.  For, YbRh$_2$Si$_2$, $T_{0}$ is estimated as 
$T_{0}=7$~mK using the band structure calculations.~\cite{Norman,Jeong}  
There are no available data for other systems shown in Table\ \ref{Table:1} 
{for the moment}.  

\subsection{Mode-mode coupling theory for CVF}
Now, we construct a self-consistent renormalization (SCR) theory for valance fluctuations.
Although higher order terms $v_{j}$ $(j\ge 3)$ in $S[\varphi]$ are irrelevant 
as shown above, the effect of their mode couplings renormalize $\eta_{0}$, inverse 
susceptibility in the RPA, and low-$T$ physical quantities 
significantly as is well known in spin-fluctuation 
theories.~\cite{Moriya,MoriyaTakimoto,Hertz,Millis}  
To construct the best Gaussian for $S[\varphi]$, we employ the so-called Feynman's inequality 
on the free energy:~\cite{Feynman}
\begin{equation}
F\le F_{\rm eff}+T\langle S-S_{\rm eff}\rangle_{\rm eff}\equiv\tilde{F}(\eta),
\label{eq:22}
\end{equation}
where
\begin{equation}
S_{\rm eff}[\varphi]\equiv\frac{1}{2}\sum_{m}\sum_{{\bf q},{l}}
(\eta+Aq^2+C_{q}|\omega_{l}|)|\varphi_{m}({\bf q},i\omega_{l})|^{2}, 
\label{eq:23}
\end{equation}
and $\eta$ is determined to make $\tilde{F}(\eta)$ be optimum. 
By optimal condition $d\tilde{F}(\eta)/d\eta=0$, 
the self-consistent equation for $\eta$, i.e., the SCR equation, is obtained as follows: 
\begin{equation}
\eta=\eta_{0}+\frac{3v_{4}}{N_{\rm s}}T\sum_{{\bf q},l}(\eta+Aq^2+C_{q}|\omega_{l}|)^{-1}.
\label{eq:24}
\end{equation}
When the system is clean and the valence change is uniform in space, 
i.e., $C_{q}=C/q$, the SCR equation in $d=3$ 
in the $Aq_{\rm B}^{2}\lsim\eta$ regime with $q_{\rm B}$ being the momentum at the Brillouin Zone 
is given by 
\begin{equation}
y=y_{0}+\frac{3}{2}y_{1}t\left[
\frac{x_{\rm c}^3}{6y}
-\frac{1}{2y}\int_{0}^{x_{\rm c}}dx\frac{x^3}{x+\frac{t}{6y}}
\right], 
\label{eq:25}
\end{equation}
where 
$y\equiv \eta/(Aq_{\rm B}^2)$, $x\equiv q/q_{\rm B}$, 
$x_{\rm c}\equiv=q_{\rm c}/q_{\rm B}$, 
and $y_0$ parameterizes a distance from the criticality and 
$y_1$ is a dimensionless mode-coupling constant of $O(1)$.  
The solution of 
Eq.\ (\ref{eq:25}) is quite different from that of ordinary SCR equation 
for spin fluctutions~\cite{Moriya} 
because of extreme smallness of $A$ in eq.\ (\ref{eq:19}). 

\begin{figure}
\begin{center}
\includegraphics[width=0.75\linewidth]{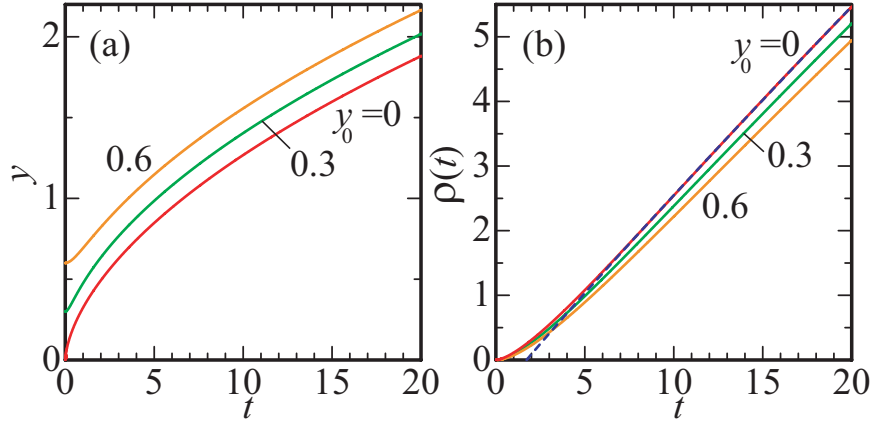} 
\caption{(Color online) 
(a) Numerical solutions of eq.\ (\ref{eq:1}) 
for $y_{0}=0.0$ (at QCP), 0.3, and 0.6 at $y_{1}=1$ and $x_{\rm c}=1$. 
(b) Electrical resistivity $\rho(T)$ calculated by using $y(t)$ in (a). 
Dashed line represents the linear-$t$ fit. 
}
\label{Fig:11}
\end{center}
\end{figure}

In the $y\gg t$ limit at QCEP with $y_{0}=0$,an analytic solution of eq.\ (\ref{eq:25}) 
is obtained as $\chi_{\rm v}(0,0)=y^{-1}\sim t^{-2/3}$ for both the clean $(z=3)$ system and 
dirty $(z=2)$ system.  
This indicates that the valence susceptibility shows unconventional criticality 
\begin{equation}
\chi_{\rm v}({\bf 0},0)=\eta^{-1}\propto t^{-2/3}. 
\label{eq:26}
\end{equation}
Figure \ref{Fig:11}(a) shows numerical solutions of Eq.\ (\ref{eq:25}) for a series of value $y_{0}$'s.  
As discussed above, the region of $t\equiv T/T_{0}$ shown in Fig.\ \ref{Fig:11} 
corresponds to that of $T\ll T_{\rm F}\sim O(D)$, so that a wide range of $t=T/T_{0}$ 
is shown in the plot even though near the criticality $y\ll 1$.  The least square fit 
of the data for $5\le t \le 100$ gives $y\propto t^{0.551}$.  If we express the inverse 
susceptibility $y$ as $y\propto t^{-\zeta}$, the exponent $\zeta$ has a temperature dependence and 
$0.5\lsim\zeta\lsim0.7$ depending on the temperature range. 

On the other hand, in the region of extremely low temperatures, $t=T/T_{0}\ll 1$, the solution 
is given by the conventional one with 
$d=3$ and $z=3$, i.e., $y(t)\propto t^{4/3}$, coinciding with that in three dimensional 
ferromagnetic QCP.  In the dirty system, with $d=3$ and $z=2$, the asymptotic form is given as 
$y(t)\propto t^{3/2}$, coinciding with that in three dimensional AF QCP.

\subsection{Critical exponents of physical quantities}
\begin{figure}
\begin{center}
\includegraphics[width=0.75\linewidth]{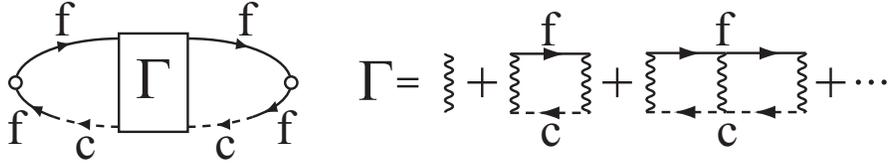} 
\caption{ 
Feynman diagrams for dynamical valence susceptibility and 
dynamical spin susceptibility for f electrons. 
Solid lines and dashed lines represent the f- and conduction-electron 
Green functions, $G^{\rm ff}_{0}$ and $G^{\rm cc}_{0}$, respectively.
Wiggly lines represent $U_{\rm fc}$. 
}
\label{Fig:12}
\end{center}
\end{figure}

The next problem is how this new type of criticality is manifested in physical quantities 
listed in Table \ref{Table:1}.  It is important to note that the valence fluctuation 
propagator $\chi_{\rm v}({\bf q},i\omega_{l})$ is qualitatively given by that of 
RPA as discussed above. 
A crucial consequence of this fact is that the dynamical $f$-spin susceptibility 
\begin{equation}
\chi_{\rm f}^{+-}({\bf q},i\omega_{l})
\equiv \int_{0}^{\beta}d\tau\langle T_{\tau}
S_{\rm f}^{+}({\bf q},\tau)S_{\rm f}^{-}({\bf -q},0)\rangle
{\rm e}^{i\omega_{l}\tau}
\label{eq:27}
\end{equation}
has the same structure as $\chi_{\rm v}$ in the RPA framework as shown in Fig.\ \ref{Fig:12}. 
At the QCEP of the valence transition, the valence susceptibility 
$\chi_{\rm v}({\bf 0},0)$ diverges. The common structure indicates that 
$\chi_{\rm f}^{+-}({\bf 0},0)$ also diverges at the QCEP. 
Then, the singularity in the  uniform spin susceptibility $\chi(T)$ is given by 
\begin{equation}
\chi\approx\chi^{\rm f}_{\rm s}\approx\frac{3}{2}\mu_{\rm B}^{2}g_{\rm f}^{2}\chi^{+-}_{\rm f}({\bf 0},0)
\propto \chi_{\rm v}(0,0),
\label{eq:28}
\end{equation}
where $\chi^{\rm f}_{\rm s}$ is the uniform $f$-spin susceptibility, 
$\mu_{\rm B}$ the Bohr magneton, and $g_{\rm f}$ Lande's g factor for $f$ electrons. 
This gives a qualitative explanation for the fact that the uniform spin susceptibility 
diverges at the QCEP of valence transition under the magnetic field, 
which was verified by the slave-boson mean-field theory applied to the extended PAM, 
Eq.\ (\ref{eq:1}).~\cite{WTMF1,WTMF2} 
Numerical calculations for the model (\ref{eq:1}) 
in $d=1$ by the DMRG~\cite{WTMF1} and in $d=\infty$ by the DMFT~\cite{Sugibayashi} 
also show the simultaneous divergence of $\chi_{\rm v}$ and uniform spin susceptibility 
under the magnetic field, again reinforcing the above argument based on RPA. 

Therefore, the uniform magnetic susceptibility $\chi(T)$ is proportional to the valence 
susceptibility $\chi_{\rm v}({\bf 0},0)$ and exhibits the same critical behavior as 
$\chi(T)\propto t^{-\zeta}$ at QCEP of valence transition.    
The spin-lattice relaxation rate $1/{(T_{1}T)}$ has the same singularity 
{in the limit} $T\to 0$ as the uniform 
susceptibility $\chi(T)$ in the case of $d=3$ and $z=3$: i.e., $1/T_{1}T\propto t^{-\zeta}$. 
Therefore, for the region $y\gg t$, $\chi(t)\sim t^{-2/3}$ and $(T_{1}T)^{-1}\sim t^{-2/3}$. 
When $T$ is decreased down to $T\sim T_{0}$, $y$ in eq.\ (\ref{eq:25}) is evaluated as 
$y\sim t^{0.5}$ by the least square fit of the numerical solution. 
Hence, depending on the flatness of critical valence fluctuation mode and 
measured temperature range, $\chi(T)\sim t^{-\zeta}$ and $(T_1T)^{-1}\sim t^{-\zeta}$ 
with $0.5\lsim \zeta \lsim 0.7$ are predicted as shown in Table\ \ref{Table:1} 
in which good agreement with experiments is manifested.   

The electrical resistivity $\rho(T)$ is calculated following a procedure used in the case of 
critical spin fluctuation as follows:~\cite{Ueda1975}
\begin{equation}
\rho(T)\propto\frac{1}{T}\int_{-\infty}^{\infty}d\omega
\omega n(\omega)[n(\omega)+1]\int_{0}^{q_{\rm c}}dqq^3{\rm Im}\chi_{\rm v}^{\rm R}(q,\omega),
\label{eq:29}
\end{equation}
where $n(\omega)=1/({\rm e}^{\beta\omega}-1)$ is the Bose distribution function, and 
$
\chi_{\rm v}^{\rm R}(q,\omega)=(\eta+Aq^2-iC_{q}\omega)^{-1}
$, 
the retarded valence susceptibility.  As for $\eta=y(t)(Aq_{\rm B}^{2})$, $y(t)$ shown in 
Fig.\ \ref{Fig:11}(a) is used for the clean system $C_{q}=C/q$.  The temperature dependence 
is shown in Fig.\ \ref{Fig:11}(b) where the normalization constant is taken as unity. 
In the region $y\gsim t$ (but $T\ll T_{\rm K}$), $\rho(t)\propto t$.  This behavior 
arises from the high-temperature limit of Bose distribution function, indicating that 
the system is described as if it is in the classical regime, because the system is 
in the high-$T$ regime in the scaled temperature $t\equiv T/T_{0}\gg 1$, in spite of 
$T\ll T_{\rm K}$.  The emergence of $\rho(t)\propto t$ behavior can be understood from 
the locality of valence fluctuations: 
In the system with an extremely small coefficient $A$ 
the dynamical exponent is regarded as almost $z=\infty$ when we write $C_{q}$ in a general 
form as $C_{q}=C/q^{z-2}$.  If we use this expression ($A=0$ corresponding to $z=\infty$) 
in $\chi_{\rm v}^{\rm R}(q,\omega)$ in the calculation of $\rho(T)$ , we easily obtain 
$\rho(T)\propto T$ toward $T\to 0$~K. 
This result indicates that the locality of valence fluctuations causes the $T$-linear resistivity. 
Indeed, the emergence of $\rho(T)\propto T$ by valence fluctuations was shown theoretically 
on the basis of the valence susceptibility $\chi_{\rm v}$ which has an approximated form for 
$z=\infty$ in Ref.\ \citen{Holmes}.  

On the other hand, in the region $t=T/T_{0}\ll 1$, the resistivity behaves as 
$\rho\sim t^{5/3}$ in the clean system and $\rho\sim t^{3/2}$ in the dirty system 
in the $t \to 0$ limit $(T \ll T_{0}\ll T_{\rm K})$. Therefore, the temperature 
dependence is expected to crossover at $T=T_{0}$.  Indeed, such a crossover has 
been observed in $\beta$-YbAlB$_4$ as shown in Table\ \ref{Table:1}.   

The specific heat $C(T)$ is estimated through the effect of self-energy of quasiparticles due to 
an exchange of valence-fluctuation modes given by (\ref{eq:19}).~\cite{Holmes,WM2006} 
A numerical solution of the self-energy gives a logarithmic temperature dependence in the 
specific-heat coefficient $C/T\propto -{\log}t$ in a certain temperature range 
$T\ll T_{0}$.  The logarithmic $t$ dependence can remain even in a range $T>T_{0}$, 
where the uniform magnetic susceptibility and the resistivity behave as $\chi(t)\sim t^{-\zeta}$ 
with $0.5\lsim \zeta \lsim 0.7$ and $\rho(t)\sim t$, respectively.  
On the other hand, it is also possible 
that{, in the case of the local limit $A\approx 0$, the power law} behavior 
$C/t\propto \chi_{\rm v}(0,0)\propto t^{-\zeta}$ appears~\cite{Holmes} before the conventional 
logarithmic behavior at high temperature region $T\lsim T_{\rm K}$, which is usually 
observed in heavy fermion metal such as CeCu$_6$,~\cite{Satoh} set{s} in.  

In conclusion of this section, 
when experimentally accessible lowest temperature is larger than $T_{0}$, 
unconventional criticality dominates all the physical quantities down to 
the lowest temperature, reproducing the unconventional criticality summarized 
in Table\ \ref{Table:1}. 

\section{Perspective}
The results on unconventional criticality due to {the} CVF succeeded in 
explaining existing experimental aspects coherently as discussed in \S3.  
On the other hand, the absolute value of change in the valence predicted 
by the theory on the extended PAM is larger than that observed in experiments 
{in general}. 
This originates from a{n inevitable} problem of the Anderson model.  
In the PAM, original one or extended one, the conduction electron{s} 
which hybridize with localized $f$ electron have the same local symmetry, 
namely the same CEF symmetry, around the Ce- or Yb-site.  
Therefore, when one measures the valence, say by Ce L$_3$ edge absorption, 
electrons with {the} local CEF symmetry would be counted together.  
Namely, a part of ``conduction electrons" (in the Anderson model) would be 
regarded as {electrons with $f$-symmetry}, or the $f$-electron state measured 
by X-ray is the hybridized object of $f$- and conduction electrons {on ligands}.  
Conduction electrons with a certain CEF symmetry at one site can 
mix with component of conduction electron with another CEF symmetry at 
different (say adjacent) sites.  This kind of effect is out of scope of our 
extended PAM. Therefore, the decrement of valence measured by experiments 
would be far smaller than that predicted by theory based on our extended PAM. 
A more realistic model including such an effect is desired.  
Nevertheless, the model Hamiltonian, (\ref{eq:1}), is useful as 
the ``fixed-point" model Hamiltonian which describes the critical behaviors associated with 
the critical valence transition.   

{A related problem is how to take into account the effect of $f$-electron state 
in the excited CEF levels.} 
In {our model,} (\ref{eq:1}), we have neglected effects of excited
CEF levels.  For the moment, it is not so clear whether those CEF states give an
essential effect {in the case of} a realistic CEF level scheme 
{measured by the 
neutron scattering experiment,~\cite{Horn}} while the effect of
charge transfer of $f$ electrons between ground and excited CEF levels has been 
discussed by Hattori~\cite{Hattori} 
{on the basis of a CEF level scheme which is practically different from 
the observed one}.  In any case, such effects certainly 
deserve further investigations.

Recently, a series of anomalies have been observed in transport properties 
of CeCu$_2$Si$_2$~\cite{Seyfarth,Araki} and $\beta$-YbAlB$_4$~\cite{Farrell}. 
Quite recently, it turned out that the CVF can give rise to an effect 
in the Hall conductivity and the Hall coefficient.
It is also expected that the Seebeck effect and the Nernst effect are 
greatly influenced by the CVF.

\section*{Acknowledgments}
We are grateful to J. Flouquet, A. T. Holmes, M. Imada, D. Jaccard, H. Mebashi, O. Narikiyo, 
Y. Onishi, T. Sugibayashi, and A. Tsuruta for collaborations on which the present article is based.   
This work was supported by JSPS KAKENHI Grant Number 25400369 and 24540378.


\begin{thebibliography}{99}

\bibitem{Steglich1}
F. Steglich, J. Aarts, C. D. Bredl, W. Lieke, D. Meschede, and W. Franz, and 
H. Sch\"afer: Phys. Rev. Lett. {\bf 43} (1979) 1892.
\bibitem{Stockert}
O. Stockert, J. Arndt, E. Faulhaber, C. Geibel, H. S. Jeevan, S. Kirchner, M. Loewenhaupt, 
K. Schmalzl, W. Schmidt, Q. Si,  and F. Steglich: 
Nature Physics {\bf 7} (2011) 119. 
\bibitem{Grosche}
F. M. Grosche, S. R. Julian, N. D. Mathur, and G. G. Lonzarich: 
Physica B {\bf 223/224} (1996) 50.
\bibitem{Mathur}  
N. D. Mathur, F. M. Grosche, S. R. Julian, I. R. Walker, D. M. Freye, 
R. K. W. Haselwimmer, and G. G. Lonzarich: Nature {\bf 394} (1998) 39. 
\bibitem{Movshovich} 
R. Movshovich, T. Graf, D. Mandrus, J. D. Thompson, J. L. Smith, and Z. Fisk:  
Phys. Rev. B {\bf 53} (1996) 8241.
\bibitem{MSRV}
K. Miyake, S. Schmitt-Rink, and C. M. Varma: Phys. Rev. B {\bf 34} (1986) 6554.
\bibitem{SLH}
D. J. Scalapino, E. Loh, Jr., and J. E. Hirsch: 
Phys. Rev. B {\bf 34} (1986) 8190.
\bibitem{Monthoux}
P. Monthoux and G. G. Lonzarich: Phys. Rev. B {\bf 63} (2001) 054529.
\bibitem{MoriyaUeda}
T. Moriya and K. Ueda: Adv. Phys. {\bf 49} (2000) 555. 
\bibitem{Bellarbi}
B. Bellarbi, A. Benoit, D. Jaccard, J. M. Mignot, and H. F. Braun: 
Phys. Rev. B {\bf 30} (1984) 1182.
\bibitem{Razafi}
H. Razafimandimby, P. Fulde, and J. Keller: Z. Phys. {\bf 54} (1984) 111.
\bibitem{Jaccard}
D. Jaccard, H. Wilhelm, K. Alami-Yadri, and E. Vargoz: Physica B {\bf 259-261} (1999) 1.
\bibitem{MNO}
K. Miyake, O. Narikiyo, and Y. Onishi: Physica B {\bf 259-261} (1999) 676. 
\bibitem{Onishi1}
Y. Onishi and K. Miyake: J. Phys. Soc. Jpn. {\bf 69} (2000) 3955. 
\bibitem{Onishi2}
Y. Onishi and K. Miyake: Physica B {\bf 281\&282} (2000) 191.  
\bibitem{MM}
K. Miyake and H. Maebashi: J. Phys. Soc. Jpn. {\bf 71} (2002) 1007.
\bibitem{WIM}
S. Watanabe, M. Imada, and K. Miyake: J. Phys. Soc. Jpn. {\bf 75} (2006) 043710.
\bibitem{Holmes}
A. T. Holmes, D. Jaccard, and K. Miyake: Phys. Rev. B {\bf 69} (2004) 024508. 
\bibitem{Yuan}
H. Q. Yuan, F. M. Grosche, M. Deppe, C. Geibel, G. Sparn, and F. Steglich: 
Science {\bf 302} (2003) 2104. 
\bibitem{Doniach}
S. Doniach: Physica B {\bf 91} (1977) 231.
\bibitem{AlphaGamma}
See for example, D. C. Koskenmaki and K. A. Geschneidner: 
{\it Handbook on the Physics and Chemistry of the Rare 
Earths}, edited by K. A. Qschneidner and L. Eyring 
(North-Holland, Amsterdam, 1978), Vol. 1, Chap. 4.
\bibitem{Coqblin} 
B. Coqblin and A. Blandin: Adv. Phys. {\bf 17} (1968) 281. 
\bibitem{Allen}
J. W. Allen and R. M. Martin: Phys. Rev. Lett. {\bf 49} (1982) 1106.  
\bibitem{Dzero}
M. Dzero, M. R. Norman, I. Paul, C. P\'epin, and J. Schmalian: 
Phys. Rev. Lett. {\bf 97} (2006) 185701.  
\bibitem{FalicovKimball}
L. M. Falicov and J. C. Kimball: Phys. Rev. Lett. {\bf 22} (1969) 997.
\bibitem{Varma1}
C. M. Varma: Rev. Mod. Phys. {\bf 48} (1976) 219.
\bibitem{HewsonRiseborough}
A. C. Hewson and P. S. Riseborough: Solid State Commun. {\bf 22} (1977) 379.
\bibitem{Schlottmann}
P. Schlottmann: Phys. Rev. B {\bf 22} (1980) 613. 
\bibitem{CostiHewson}
T. A. Costi and A. C. Hewson: Physica C {\bf 185-189} (1991) 2649.
\bibitem{TakayamaSakai}
R. Takayama and O. Sakai: J. Phys. Soc. Jpn. {\bf 66} (1997) 1512.
\bibitem{Perakis}
I. E. Perakis and C. M. Varma: Phys. Rev. B {\bf 49} (1994) 9041.
\bibitem{Khomskii}
D. I. Khomskii and A. N. Kocharjan: Solid State Commun. {\bf 18} (1976) 985.
\bibitem{Saiga}
Y. Saiga, T. Sugibayashi, and D. S. Hirashima: 
J. Phys. Soc. Jpn. {\bf 77} (2008) 114710.
\bibitem{Sugibayashi}
T. Sugibayashi, A. Tsuruta, and K. Miyake: Physica C {\bf 470} (2010) S550. 
\bibitem{Kubo}
K. Kubo: J. Phys. Soc. Jpn. {\bf 80} (2011) 114711. 
\bibitem{Hagymasi}
I. Hagymasi, K. Itai, and J. Solyom: Acta Phys. Pol. A {\bf 121} (2012) 1070. 
\bibitem{Bauer}
E. Bauer, R. Hauser, L. Keller, P. Fischer, O. Trovarelli, J. G. Sereni, 
J. J. Rieger, and G. R. Stewart: Phys. Rev. B {\bf 56} (1997) 711.
\bibitem{Seuring1}
C. Seuring, K. Heuser, E.-W Scheidt, T. Schreiner, E. Bauer, and G. R. Stewart: 
Physica B {\bf 281-282} (2000) 374. 
\bibitem{Steglich2}
O. Trovarelli, C. Geibel, S. Mederle, C. Langhammer, 
F.M. Grosche, P. Gegenwart, M. Lang, G. Sparn, and F. Steglich: 
Phys. Rev. Lett. {\bf 85} (2000) 626.
\bibitem{Ishida}
K. Ishida, K. Okamoto, Y. Kawasaki, Y. Kitaoka, O. Trovarelli, C. Geibel, and F. Steglich:  
Phys. Rev. Lett. {\bf 89} (2002) 107202. 
\bibitem{Loehneysen}
H. v. L\"ohneysen, A. Rosch, M. Vojta, and P. W\"olfle: 
Rev. Mode. Phys. {\bf 79} (2007) 1015. 
\bibitem{Nakatsuji}
S. Nakatsuji, K. Kuga, Y. Machida, T. Tayama, T. Sakakibara, Y. Karaki, H. Ishimoto, 
S. Yonezawa, Y. Maeno, E. Pearson, G. G. Lonzarich, L. Balicas, H. Lee, and Z. Fisk: 
Nature Phys. {\bf 4} (2008) 603.  
\bibitem{Matsumoto}
Y. Matsumoto, S. Nakatsuji, K. Kuga, Y. Karaki, N. Horie, Y. Shimura, T. Sakakibara, 
A. H. Nevidomskyy, and P. Coleman: 
Science {\bf 331} (2011) 316.
\bibitem{Deguchi}
K. Deguchi, S. Matsukawa, N. K. Sato, T. Hattori, K. Ishida, H. Takakura, and T. Ishimasa: 
Nature Mat. {\bf 11} (2012) 1013. 
\bibitem{Moriya}
T. Moriya : {\it Spin Fluctuations in Itinerant Electron Magnetism} 
(Springer, Berlin, 1985).
\bibitem{MoriyaTakimoto} 
T. Moriya and T. Takimoto: J. Phys. Soc. Jpn. {\bf 64} (1995) 960.
\bibitem{Hertz}
J. A. Hertz: Phys. Rev. B {\bf 14} (1976) 1165.
\bibitem{Millis}
A. J. Millis: Phys. Rev. B {\bf 48} (1993) 7183. 
\bibitem{Hatatani}
M. Hatatani, O. Narikiyo and K. MiyakeFJ. Phys. Soc. Jpn. {\bf 67} (1998) 4002; M.
Hatatani, PhD Thesis, 2000, Graduate School of Engineering Science, Osaka University.
\bibitem{WM:PRL}
S. Watanabe and K. Miyake, Phys. Rev. Lett. {\bf 105} (2010) 186403.
\bibitem{WM:JPCM}
S. Watanabe and K. Miyake, J. Phys.: Condens. Matter {\bf 24} (2012) 294208. 
\bibitem{Si}
Q. Si, S. Rabello, K. Ingersent, and J. L. Smith: Nature {\bf 413} (2001) 804.
\bibitem{Coleman}
P. Coleman C. P\'epin, Q. Si, and R. Ramazashvili: J. Phys.: Condens. Matter {\bf 13} 
(2001) R723. 
\bibitem{Si2}
Q. Si: Physica B {\bf 378-380} (2006) 23.
\bibitem{Misawa}
T. Misawa, Y. Yamaji, and M. Imada: J. Phys. Soc. Jpn. {\bf 78} (2009) 084707. 
\bibitem{Coleman2}
A. Ramires, P. Coleman, A. H. Nevidomskyy, and A. M. Tsvelik: 
Phys. Rev. Lett. {\bf 109} (2012) 176404.
\bibitem{RiceUeda}
T. M. Rice and K. Ueda: Phys. Rev. B {\bf 34} (1986) 6420.  
\bibitem{Shiba}
H. Shiba: J. Phys. Soc. Jpn. {\bf 55} (1986) 2765.
\bibitem{rf:KW}
K. Kadowaki and S. B. Woods: Solid State Commun. {\bf 58} (1986) 507.
\bibitem{rf:MMV}
K. Miyake, T. Matsuura, and C. M. Varma: Solid State Commun. {\bf 71} (1989) 1149. 
\bibitem{MN}
K. Miyake and O. Narikiyo: J. Phys. Soc. Jpn. {\bf 71} (2002) 867. 
\bibitem{Miyake:JPCM}
K. Miyake: J. Phys.: Condens. Matter {\bf 19} (2007) 125201. 
\bibitem{Fujiwara}
K. Fujiwara, Y. Hata, K. Kobayashi, K. Miyoshi, J. Takeuchi, Y. Shimaoka, H. Kotegawa, 
T. C. Kobayashi, C. Geibel, and F. Steglich: J. Phys. Soc. Jpn. {\bf 77} (2008) 123711. 
\bibitem{Fujiwara2}
K. Fujiwara, M. Iwata, Y. Okazaki, Y. Ikeda, S. Araki, T. C. Kobayashi, K. Murata, C. Geibel, 
and F. Steglich: J. Phys.: Conf. Ser. {\bf 391} (2012) 012012. 
\bibitem{Kobayashi}
T. C. Kobayashi, K. Fujiwara, K. Takeda, H. Harima, Y. Ikeda, T. Adachi, Y. Ohishi, C. Geibel, 
and F. Steglich: J. Phys. Soc. Jpn. {\bf 82} (2013) 114701.
\bibitem{Onodera}
A. Onodera, S. Tsuduki, Y. Ohishi, T. Watanuki, K. Ishida, Y. Kitaoka, and Y. \=Onuki: 
Solid State Commun. {\bf 123} (2002) 113. 
\bibitem{Rueff}
J.-P. Rueff, S. Raymond, M. Taguchi, M. Sikora, J.-P. Iti\'e, F. Baudelet, D. Braithwaite, G. Knebel, 
and D. Jaccard: Phys. Rev. Lett. {\bf 106} (2011) 186405. 
\bibitem{Knebel}
G. Knebel, D. Aoki, J.-P. Brison, and J. Flouquet: J. Phys. Soc. Jpn. {\bf 77} (2008) 114704.  
\bibitem{Park}
T. Park, V. A. Sidorov, F. Ronning, J.-X. Zhu, Y. Tokiwa, H. Lee, E. D. Bauer, 
R. Movshovich, J. L. Sarrao, and J. D. Thompson: Nature {\bf 456} (2008) 366. 
\bibitem{Chen}
G. F. Chen, K. Matsubayashi, S. Ban, K. Deguchi, and N. K. Sato: 
Phys. Rev. Lett. {\bf 97} (2006) 017005.
\bibitem{Shishido}
H. Shishido, R. Settai, H. Harima, and Y. \=Onuki: 
J. Phys. Soc. Jpn. {\bf 74} (2005) 1103. 
\bibitem{WM:JPSJ}
S. Watanabe and K. Miyake: J. Phys. Soc. Jpn. {\bf 79} (2010) 033707. 
\bibitem{WM:JPCM2}
S. Watanabe and K. Miyake, J. Phys.: Condens. Matter {\bf 23} (2011) 094217. 
\bibitem{Comment}
The result of the first-order AF transition is obtained for the band structure for 
$\beta_2$ branch in CeRhIn$_5$ by the slave-boson mean-field calculation.  
It is noted that the order of the AF transition (as a consequence of $P_{\rm c}\simeq P_{\rm v}$) 
is considered to depend on the detail of the parameter set of the extended PAM and calculation 
scheme.    
\bibitem{Muramatsu}
T. Muramatsu, N. Tateiwa, T. C. Kobayashi, K. Shimizu, K. Amaya, D. Aoki, H. Shishido, 
Y. Haga, and Y. \=Onuki: J. Phys. Soc. Jpn. {\bf 70} (2001) 3362.
\bibitem{Landau}
L. D. Landau: Zh. Exp. Teor. Fiz. {\bf 7} (1937) 19; ibid. 627; 
L. D. Landau: {\it Collected papers of L. D. Landau}, (Pergamon Press, Oxford, 1965) p. 193; 
L. D. Landau and E. M. Lifshitz: {\it Statistical Physics}, (Pergamon Press, London-Paris,
1958) 1st edition, \S116 and \S135.
\bibitem{Norman}
M. R. Norman: Phys. Rev. B {\bf 71} (2005) 22045R. 
\bibitem{Jeong}
T. Jeong and W. E. Pickett: J. Phys.: Condens. Matter {\bf 18} (2006) 6289.
\bibitem{Feynman}
R. Feynman: {\it Statistical Mechanics: A Set Of Lectures} (Advanced Books Classics). 
\bibitem{WTMF1}
S. Watanabe, A. Tsuruta, K. Miyake and J. Flouquet: Phys. Rev. Lett. {\bf 100} (2008) 236401.
\bibitem{WTMF2}
S. Watanabe, A. Tsuruta, K. Miyake and J. Flouquet: J. Phys. Soc. Jpn. {\bf 78} (2009) 104706.
\bibitem{Ueda1975}
K. Ueda and T. Moriya: J. Phys. Soc. Jpn. {\bf 39} (1975) 605.
\bibitem{WM2006}
S. Watanabe and K. Miyake: arXiv:0906.3986
\bibitem{Satoh}
K. Satoh, T. Fujita, Y. Maeno, Y. \=Onuki, and T. Komatsubara: 
J. Phys. Soc. Jpn. {\bf 58} (1989) 1012.  
\bibitem{Horn}
S. Horn, E. Holland-Moritz, M. Loewenhaupt, F. Steglich, H. Scheuer, A. Benoit, and J. Flouquet:  
Phys. Rev. B {\bf 23} (1981) 3171. 
\bibitem{Hattori}
K. Hattori: J. Phys. Soc. Jpn. {\bf 79} (2010) 114717.
\bibitem{Seyfarth}
G. Seyfarth, A.-S. R\"uetschi, K. Sengupta, A. Georges, D. Jaccard, 
S. Watanabe, and K. Miyake: Phys. Rev. B {\bf 85} (2012) 205105.
\bibitem{Araki}
S. Araki, Y. Shiroyama, Y. Ikeda, T. C. Kobayashi, S. Seiro, C. Geibel, and F. Steglich: 
J. Phys. Soc. Jpn. {\bf 80} (2011) SA061; private communications.
\bibitem{Farrell}
E. C. T. OfFarrell, Y. Matsumoto, and S. Nakatsuji: Phys. Rev. Lett. {\bf 109} (2012) 176405.



\end{thebibliography}
\end{document}